\documentclass[useAMS,usenatbib]{mn2e}

\usepackage[dvips]{graphicx}
\usepackage{epsfig,natbib}
\usepackage{mathptm,amsmath,amssymb}

\def\reff@jnl#1{{\rm#1\/}}
\def\aj{\reff@jnl{AJ}}                 % Astronomical Journal
\def\araa{\reff@jnl{ARA\&A}}           % Annual Review of Astron and Astrophys
\def\apj{\reff@jnl{ApJ}}               % Astrophysical Journal
\def\apjl{\reff@jnl{ApJ}}              % Astrophysical Journal, Letters
\def\apjs{\reff@jnl{ApJS}}             % Astrophysical Journal, Supplement
\def\ao{\reff@jnl{Appl.Optics}}        % Applied Optics
\def\apss{\reff@jnl{Ap\&SS}}           % Astrophysics and Space Science
\def\aap{\reff@jnl{A\&A}}              % Astronomy and Astrophysics
\def\aapr{\reff@jnl{A\&A~Rev.}}        % Astronomy and Astrophysics Reviews
\def\aaps{\reff@jnl{A\&AS}}            % Astronomy and Astrophysics, Supplement
\def\azh{\reff@jnl{AZh}}               % Astronomicheskii Zhurnal
\def\baas{\reff@jnl{BAAS}}             % Bulletin of the AAS
\def\jrasc{\reff@jnl{JRASC}}           % Journal of the RAS of Canada
\def\memras{\reff@jnl{MmRAS}}          % Memoirs of the RAS
\def\mnras{\reff@jnl{MNRAS}}           % Monthly Notices of the RAS
\def\pra{\reff@jnl{Phys.Rev.A}}        % Physical Review A: General Physics
\def\prb{\reff@jnl{Phys.Rev.B}}        % Physical Review B: Solid State
\def\prc{\reff@jnl{Phys.Rev.C}}        % Physical Review C
\def\prd{\reff@jnl{Phys.Rev.D}}        % Physical Review D
\def\prl{\reff@jnl{Phys.Rev.Lett}}     % Physical Review Letters
\def\pasp{\reff@jnl{PASP}}             % Publications of the ASP
\def\pasj{\reff@jnl{PASJ}}             % Publications of the ASJ
\def\qjras{\reff@jnl{QJRAS}}           % Quarterly Journal of the RAS
\def\skytel{\reff@jnl{S\&T}}           % Sky and Telescope
\def\solphys{\reff@jnl{Solar~Phys.}}   % Solar Physics
\def\sovast{\reff@jnl{Soviet~Ast.}}    % Soviet Astronomy
\def\ssr{\reff@jnl{Space~Sci.Rev.}}    % Space Science Reviews
\def\zap{\reff@jnl{ZAp}}               % Zeitschrift fuer Astrophysik
\def\nat{\reff@jnl{Nature}}            % Nature
\def\tfrac#1#2{{\textstyle\frac{#1}{#2}}}
\def\vect#1{{\mathbfit{#1}}}

\def\be{\begin{equation}}
\def\ee{\end{equation}}
\def\ber{\begin{eqnarray}}
\def\eer{\end{eqnarray}}

\title[Fast cosmological parameter estimation]
{Fast cosmological parameter estimation using neural networks}
\author[T.~Auld et al.]
{T.~Auld, M.~Bridges, M.P.~Hobson and S.F.~Gull\\
Astrophysics Group, Cavendish Laboratory, Magingley Road,
Cambridge CB3 0HE, UK}

\date{Accepted ---. Received ---; in original form \today}

\pagerange{\pageref{firstpage}--\pageref{lastpage}} \pubyear{2004}

\begin{document}

\label{firstpage}

\maketitle

\begin{abstract}
\noindent We present a method for accelerating the calculation of
CMB power spectra, matter power spectra and likelihood functions
for use in cosmological parameter estimation. The algorithm,
called {\sc CosmoNet}, is based on training a multilayer perceptron
neural network and shares all the advantages of the recently
released {\sc Pico} algorithm of Fendt \& Wandelt, but has several
additional benefits in terms of simplicity, computational speed,
memory requirements and ease of training.  We demonstrate the
capabilities of {\sc CosmoNet} by computing CMB power spectra over a box
in the parameter space of flat $\Lambda$CDM models containing the
$3\sigma$ WMAP1 confidence region. We also use {\sc CosmoNet} to compute
the WMAP3 likelihood for flat $\Lambda$CDM models and show that
marginalised posteriors on parameters derived are very similar to
those obtained using {\sc CAMB} and the WMAP3 code. We find that the
average error in the power spectra is typically $2-3\%$ of cosmic
variance, and that {\sc CosmoNet} is $\sim 7 \times 10^4$ faster than
{\sc CAMB} (for flat models) and $\sim 6 \times 10^6$ times faster than
the official WMAP3 likelihood code. {\sc CosmoNet} and an interface to
{\sc CosmoMC} are publically available at {\tt
www.mrao.cam.ac.uk/software/cosmonet}.
\end{abstract}


%----------------------------------------------------------------------

\section{Introduction}
\label{sec:intro}

In the analysis of increasingly high-precision data sets, it is now
common practice in cosmology to constrain cosmological parameters
using sampling based methods, most notably Markov chain Monte Carlo
(MCMC) techniques (Christensen et al. 2001; Knox, Christensen \&
Skordis 2001; Lewis \& Bridle 2002). This approach typically requires
one to calculate theoretical CMB power spectra (i.e. some subset of
the TT, TE, EE and BB $C_\ell$ spectra) and/or the matter power
spectrum $P(k)$ at a large number of points (typically $\sim 10^5$ or
more) in the cosmological parameter space. In addition, one must also
evaluate at each point the corresponding (combined) likelihood
function for the data set(s) under consideration. As a result, the
process can be computational very demanding.

The purist would calculate the required power spectra at each point
using codes such as {\sc CMBfast} (Seljak \& Zaldarriaga 1996) or {\sc CAMB}
(Lewis, Challinor \& Lasenby 2000), which typically require around 10
secs for spatially-flat models
and 50 secs for non-flat models. This approach is therefore
computationally demanding, but does have the advantage that it is
simple to generalise if one wishes to include new physics or change
the form of the initial power spectra. MCMC parameter estimation codes
such as {\sc CosmoMC} (Lewis \& Bridle 2002) attempt to decrease the overall
computational burden by dividing the cosmological parameter space into
`fast' parameters (governing the initial primoridal power spectra of
scalar and tensor perturbations) and `slow' parameters (governing the
perturbation evolution) and making judicious proposals for how the
chain is propagated in parameter space.  Even with this technique,
however, the total computational cost is still usually very high.

If one is willing to forego the full calculation of the required
power spectra at each point in parameter space, there are a number of
ways in which suitably accurate spectra can be generated somewhat more
rapidly. If the cosmological parameter space of interest is
sufficiently small, then it is possible simply to create spectra for a
regular grid of models in parameter space and interpolate between them
in some way. As the number of parameters increases, however, the
computational cost of constructing the grid grows exponentially. Fast
grid generation schemes have been proposed, such as the
$\ell$-splitting scheme of Tegmark \& Zaldarriaga (2000) that exploits
analytic approximations at high-$\ell$ and insensitivity to certain
parameters at low-$\ell$.
%With this technique they were able to
%calculate a 7-dimensional grid of $C_\ell$ spectra (for TT) with many
%fewer calls to CMBfast than would be required by a brute-force method.
%Moreover, by using semi-analytic formula to correct for reionisation
%effects and scaling the scalar and tensor power spectra separately, it
%was possible to cover a 10-dimensional parameter space.
Nevertheless, the pre-compute of the grid of models remains extremely
time-consuming and such approaches become difficult to implement
accurately when second-order effects such as gravitational lensing are
important.

More extensive use of analytic and semi-analytic approximations
can reduce the required number of pre-computed models, but only at
the cost of a loss of accuracy and/or placing restrictions on the
parameters that are available as input.  Such approaches are
usually based on a relatively sparse grid of base models in the
parameter space from which the spectra of more general models are
computed rapidly on-the-fly using various (semi-)analytic
approximations.  The {\sc DASh} code of Kaplinghat, Knox \&
Skordis (2002), instead stores a sparse grid of transfer functions
(rather than $C_\ell$), uses efficient choices for grid parameters
and makes considerable use of analytic approximations. Following
$\sim 40$ hrs of computation on a typical desktop to calculate the
grid, DASh provides a speed-up factor of $\sim 30$ relative to
CMBfast in calculating a $C^{\rm TT}_\ell$, $C^{\rm TE}_\ell$ or
$C^{\rm EE}_\ell$ spectrum.  More recently, the need to
pre-compute a grid of models has been removed in the {\sc CMBwarp}
package (Jimenez et al. 2004), which builds on the method
introduced by Kosowsky, Milosavljevic \& Jimenez (2002).  In this
approach, a new set of nearly uncorrelated `physical parameters'
are introduced upon which the CMB power spectra have a simple
dependence. {\sc CMBwarp} uses a modified polynomial fit in these
parameters in which the coefficients are based on the spectra
$C^{\rm TT}_\ell$, $C^{\rm TE}_\ell$ or $C^{\rm EE}_\ell$ for just
a single fiducial model in the parameter space. Spectra for other
models can then be calculated around $\sim 3000$ times faster than
{\sc CMBfast}.  By taking the fiducial model to be the best-fit
model to the WMAP1 data, {\sc CMBwarp} gives better than 0.5 per
cent accuracy for the $C^{\rm TT}_\ell$ spectrum throughout the
entire region of parameter space lying within the WMAP1 3$\sigma$
confidence region, although the accuracy quickly reduces as one
moves further away from the fiducial model.

%It is also worth mentioning another approach due to Sandvik et
%al. (2004) that avoids the calculation of CMB power spectra
%altogether. Their {\sc CMBfit} code provides a semi-analytic fit
%directly to the WMAP1 likelihood as a function of cosmological
%parameters. Although such an approach is clearly tied to a single
%experiment (or set of experiments), it is nonetheless very
%useful. This is particularly true in analysing WMAP 3-year data
%(Hinshaw et al. 2006; Page et al. 2006; Spergel et al. 2006), for
%which calculation of the likelihood function is very time-consuming.

Although the above methods have proved extremely useful in
performing cosmological parameter estimation, they do exhibit a
number of drawbacks, as we have outlined. Most recently, this has
led Fendt \& Wandelt (2006) to propose a more flexible and robust
machine-learning approach (called {\sc Pico}) to accelerating both
power spectra and likelihood evaluations.  In this method, one
first calculates the required spectra (usually $C^{\rm TT}_\ell$,
$C^{\rm TE}_\ell$ or $C^{\rm EE}_\ell$) using {\sc CAMB} and the
corresponding likelihoods for the experiments of interest (in
particular WMAP3) at $\sim 10^4$ points chosen uniformly within a
box in parameter space that encompasses (say) the $3\sigma$
confidence region of the WMAP3 likelihood. This constitutes the
training set for the {\sc Pico} code -- note that only power
spectra values at the limited number of $\ell$-values output by
{\sc CAMB} are used (typically $50$ values for $\ell_{\rm
max}=1500$). In short, the basic algorithm used by {\sc Pico}
consists of three major parts. First, the training set is
compressed using Karhunen--Lo\`eve eigenmodes (essentially a
principal component analysis) which typically results in a
reduction in the dimensionality of the training set by a factor of
two. Second, the training set is used to divide the parameter
space into ($\sim 100$) smaller regions using a $k$-means
clustering algorithm (see e.g. MacKay 1997) with the goal that all
clusters encompass volume of parameter space over which the power
spectra vary roughly equally. Finally, a (4th order) polynomial is
fitted within each cluster (by minimising the squared error) to
provide a local interpolation of the power spectra within the
cluster as a function of cosmological parameters.  The reason for
dividing up the parameter space in the second step is that the
interpolation method used fails to model accurately the power
spectra over the entire parameter space.

The {\sc Pico} approach provides about the same speed-up in spectrum calculation
as {\sc CMBWarp} (which is an order of magnitude faster than {\sc DASh}), but is
an order of magnitude more accurate. It also has several other
important advantages. First, it is very flexible and can easily be
applied to the fast calculation of any observables relevant to a
particular data set, such as scalar, tensor and lensed power spectra,
transfer functions or even higher-order correlation functions. Second,
it allows the calculation of such observables from an arbitrary number
of cosmological models and in any range of $\ell$ (or $k$)
values. Lastly, the algorithm is sufficiently generic to allow the
direct fitting of likelihood functions, thereby incorporating the
functionality of the {\sc CMBfit} code of Sandvik et al.  This last
capability allows an additional order of magnitude speed-up in
cosmological parameter estimation beyond that resulting from faster
power spectrum calculations, and is particularly important for
experiments such as WMAP3 for which the likelihood calculation is very
expensive.

In this letter, we present an independent approach to using
machine-learning techniques for accelerating both power spectra and
likelihood evaluations. Our approach is based on training a neural
network in the form of a 3-layer perceptron. The resulting {\sc CosmoNet}
code shares all the advantages of {\sc Pico}, but we believe also has some
additional benefits in terms of simplicity, computational speed,
accuracy, memory requirements and ease of training. The letter is
organised as follows. In Section~\ref{sec:nn} we give a brief
introduction to neural networks and our training algorithm. The
resulting network output is discussed in Section~\ref{sec:interp},
where we investigate the accuracy of our approach.  The {\sc CosmoNet} code
is then used to perform a cosmological parameter estimation from WMAP3
data in Section~\ref{sec:cosmo}. Our conclusions are presented in
Section~\ref{sec:conc}.

\section{Neural network interpolation}
\label{sec:nn}

Neural networks are a methodology for computing motivated by the
parallel architecture of animal brains. They consist of a group of
interconnected processing elements called neurons that pass simple
scalar messages between them to process information. Many neural
networks provide feed-forward maps from a set of input neurons to a
set of output neurons. For an introduction to feed-forward neural
networks see Bailer-Jones et al. (2001). They are often used to provide
empirical models for processes that are too complicated to model from
theoretical principles. An astrophysical example is presented in
Vanzella et al. (2004), where photometric redshifts are predicted
in the HDF-S from an ultra deep multicolour catalogue.

\subsection{Multilayer perceptron networks}
\label{sec:mlp}

The perceptron (Rosenblatt 1958) is the simplest type of
feed-forward neuron and maps an input vector $\vect{x} \in
\Re^n$ to a scalar output $f(\vect{x};\vect{w},\theta)$ via
\be
f(\vect{x};\vect{w},\theta) = \sum_i w_{i} x_i  +  \theta,
\label{eq:perceptron}
\ee
where $\{w_{i}\}$ and $\theta$ are the parameters of the
perceptron, called the `weights' and `bias' respectively.

\begin{figure}[t]
\begin{center}
\includegraphics[width=2 in]{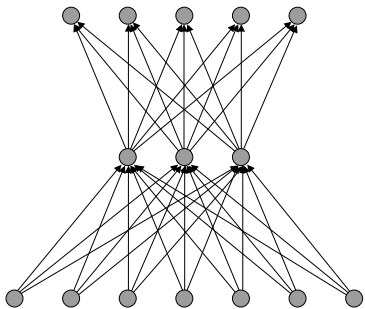}
\end{center}
\caption{\label{fig:nn} An example of a 3-layer neural network with
seven input nodes, 3 nodes in the hidden layer and five output
nodes. Each line represents one weight.}
\end{figure}

Multilayer perceptron neural networks (MLPs) are a type of
feed-forward network composed of a number of ordered layers of
perceptron neurons that pass scalar messages from one layer to the
next. In the simplest case, the network has two layers: the input
layer and and the output layer.  Each node in the output layer is a
perceptron and has an activation given by (\ref{eq:perceptron}). In
this paper, however, we will work with a 3-layer network, which
consists of an input layer, a hidden layer and an output layer, as
illustrated in Fig.~\ref{fig:nn}.
In such a network, the outputs of the nodes in the hidden and output
layers take the form
\begin{eqnarray}
\mbox{hidden layer:} & h_j=g^{(1)}(f_j^{(1)}); &
f_j^{(1)} = \sum_l w^{(1)}_{jl}x_l +
  \theta_j^{(1)}, \\
\mbox{output layer:} & y_i=g^{(2)}(f_i^{(2)}); & f_i^{(2)} =
\sum_j w^{(2)}_{ij}h_j + \theta_i^{(2)},
\end{eqnarray}
where the index $l$ runs over input nodes, $j$ runs over hidden nodes
and $i$ runs over output nodes. The functions $g^{(1)}$ and $g^{(2)}$
are called activation functions and are chosen to be bounded, smooth
and monotonic. In this letter, we use $g^{(1)}(x)=\tanh x$ and
$g^{(2)}(x)=x$, where the non-linear nature of the former is a key
ingredient in constructing a viable network.

The weights $\vect{w}$ and biases $\btheta$ are the quantities we
wish to determine, which we denote collectively by $\vect{a}$. As
these parameters vary a very wide range of non-linear mappings
between the inputs and outputs are possible.  In fact,
according to a `universal approximation theorem' (Leshno et al.
1993), a standard multilayer feed-forward network with a locally
bounded piecewise continuous activation function can approximate
any continuous function to {\it any} degree of accuracy if (and
only if) the network's activation function is not a polynomial.
This result applies when activation functions are chosen apriori
and held fixed as $\vect{a}$ varies. Accuracy increase with the
number in the hidden layer and the above theorem tells us we can
always choose sufficient hidden nodes to produce any accuracy.
Since the mapping from cosmological parameter space to the space
of CMB power spectra (and WMAP3 likelihood) is known to be
continuous, a 3-layer MLP with an appropriate choice of activation
function is an excellent candidate model for the replacement of
the forward model provided by the CAMB package (and WMAP3
likelihood code).

The activation functions act as basic building blocks of
non-linearity in a neural network model and should be as simple as
possible. Additionally, the MemSys routines used in training
(described below) require derivative information and so they
should be differentiable. The universal approximation theorem thus
motivates us to choose a monotonic (for simplicity), bounded and
differentiable function that is not a polynomial and we choose the
$tanh$ function. Of course, this could be replaced by another such
function, such as the sigmoid function, but the interpolation
results will be almost identical.

\subsection{Network training}
\label{sec:training}

Let us consider building an empirical model of the {\sc CAMB} mapping using
a 3-layer MLP as described above (a model of the WMAP3 likelihood code
can be constructed in an analogous manner). The number of nodes in the
input layer will correspond to the number of cosmological parameters,
and the number in the output layer will be the number of
uninterpolated $C_\ell$ values output by {\sc CAMB}. A set of training data
${\cal{D}} = \{\vect{x}^{(k)},\vect{t}^{(k)}\}$ is provided by {\sc CAMB} (the
precise form of which is described later) and the problem now reduces
to choosing the appropriate weights and biases of the neural network
that best fit this training data.

As the {\sc CAMB} mapping is exact, this is a deterministic problem, not a
probabilistic one. We therefore wish to choose network parameters
$\vect{a}$ that minimise the `error' term $\chi^2(\vect{a}))$ on the
training set given by
\be \chi^2(\vect{a}) = \tfrac{1}{2}\sum_k
\sum_i\left[t^{(k)}_i-y_i(\vect{x}^{(k)};\vect{a})\right]^2.
\ee
This is, however, a highly non-linear, multi-modal function in many
dimensions whose optimisation poses a non-trivial problem. Despite the
deterministic nature of the problem we use an extension of a Bayesian
method provided by the {\sc MemSys} package (Gull \& Skilling 1999).

The {\sc MemSys} algorithm considers the parameters $\vect{a}$ of the
network to be probabilistic variables with prior probability
distribution proportional to $\exp(-\alpha S(\vect{a}))$, where
$S(\vect{a})$ is the positive-negative entropy functional (Gull \&
Skilling 1999; Hobson
\& Lasenby 1998)  and $\alpha$ is considered a hyperparameter of the
prior. The variable $\alpha$ sets the scale over which variations in
$\vect{a}$ are expected, and is chosen to maximise its marginal
posterior probability. Its value is inversely proportional to the
standard deviation of the prior.  For fixed $\alpha$, the
log-posterior is thus proportional to $- \chi^2(\vect{a}) + \alpha
S(\vect{a})$.  For each choice of $\alpha$ there is a solution
$\hat{\vect{a}}$ that maximises the posterior. As $\alpha$ varies, the
set of solutions $\hat{\vect{a}}$ is called the `maximum-entropy
trajectory'. We wish to find the maximum of $-\chi^2$ which is the
solution at the end of the trajectory where $\alpha=0$. It is
difficult to recover results for $\alpha \neq \infty$ (for large
$\alpha$ the solution is found at the maximum of the prior) when
starting with a result that lies far from the trajectory. Thus for
practical purposes, it is best to start from the point on the
trajectory at $\alpha = \infty$ and iterate $\alpha$ downwards until
either a Bayesian $\alpha$ is acheived, or in our deterministic case,
$\alpha$ is sufficiently small that the posterior is dominated by $\chi^2$.

{\sc MemSys} performs the algorithm using conjugate gradients at each step
to converge to the maximum-entropy trajectory. The required matrix of
second derivatives of $\chi^2$ is approximated using vector routines
only. This avoids the need for the $O(N^3)$ operations required to
perform exact calculations, that would be impractical for large
problems. The application of {\sc MemSys} to the problem of network training
allows for the fast efficient training of relatively large network
structures on large data sets that would otherwise be difficult to
perform in a useful time-frame.  The {\sc MemSys} algorithms are described
in greater detail in (Gull \& Skilling 1999).

\section{Results}
\label{sec:interp}

We demonstrate our approach by training networks to replace the
{\sc CAMB} package for the evaluation of the CMB power spectra $C^{\rm
TT}_\ell$, $C^{\rm TE}_\ell$ or $C^{\rm EE}_\ell$ for flat
$\Lambda$CDM models within a box in parameter space that
encompasses the $3\sigma$ confidence region of the WMAP 1-year
likelihood. We also train a network to replace the WMAP 3-year
likelihood code, however for reasons to be discussed later in this section, this
interpolation was preformed over a slightly smaller region than $3 \sigma$. We
train four seperate networks: one for each CMB power spectra and one for the WMAP
 3-year likelihood. It is possible to provide all spectra and the likelihood
 from a single network, but training speed is increased by keeping
them separate.

The training data for the spectra interpolation is produced in a
similar way to that used to train the {\sc Pico} algorithm. We define
the same box as Fendt \& Wandelt in the 6-dimensional `physical
parameter' space of flat $\Lambda$CDM models, which encompasses
the $3\sigma$ confidence region of the likelihood determined from
WMAP 1-year data (Bennett et al. 2003) and other higher resolution
CMB data. This box is then sampled uniformly to select $2000$
models. The physical parameters ($\omega_{\rm b}$, $\omega_{\rm cdm}$, $\theta$,
$\tau$, $n_{\rm s}$, $A_{\rm s}$) are converted back to cosmological
parameters ($\Omega_{\rm b}$, $\Omega_{\rm cdm}$, $H_0$,
$z_{ \rm re}$, $n_{\rm s}$, $A_{\rm s}$) and used as input to CAMB to
produce the training set of CMB power spectra out to $\ell_{\rm
max} = 1500$ (which corresponds to 50 uninterpolated $C_\ell$
values for each spectrum). A further set of $10^4$ samples were
generated as testing data.

Building a training set for the likelihood was complicated by
errors in the WMAP 3-year likelihood code. Spuriously high
likelihoods were observed for some models lying outside of roughly
$2 \sigma$\footnote{An example point being: $\omega_b = 0.016048, \omega_{cdm} = 0.177486,
\theta = 1.056867, \tau = 0.501029, n_{\rm s} = 1.078956, A_{\rm s} = 3.022956$ with $\ln$-likelihood
= -5373}. These spikes in the likelihood surface prevented a
reasonable interpolation in these areas and so had to be
eliminated from the training set. In addition sampling uniformly
from a $3 \sigma$ region in 6 dimensions returned very few samples
around the maximum likelihood point --making an accurate
interpolation around the peak unworkable. To correct for both of
these problems we built our likelihood training set from $5000$
samples in parameter space drawn from a Gaussian distribution
centered on the maximum likelihood point (restricted to the box
encompassing our parameter priors). The covariance matrix of the
Gaussian was twice that of the expected variance of the
cosmological parameters and was found to provide sufficient
coverage, both for the peaks of the marginalised posteriors and their tails.

A small pre-processing step was used to make the variation in the
training data of a similar order to the non-linearity present in
the network activation functions. This involved mapping all inputs
and outputs linearly so they had zero mean and a variance of
one-half. Appropriate scaling of the data would be performed by
network training if this step were omitted, but the speed of
training is increased if the initial values of the weights are
closer to their likely optimal values. Also, for the TE and EE
spectra, a small number (2-4) of separate neural networks were
trained on separate regions of the spectra, and then combined post
training to provide a single network for each spectra. This
step was required to provide $99^{\hbox{th}}$ percentile error 
within those produced by the {\sc Pico} algorithm.

\begin{figure*}
\begin{center}
  \includegraphics[width=2.2 in]{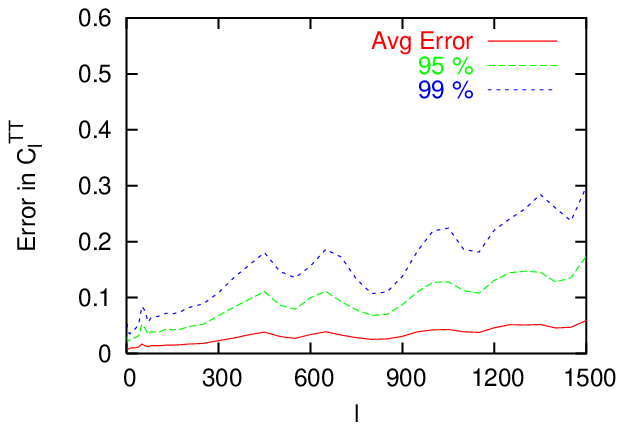}
  \includegraphics[width=2.2 in]{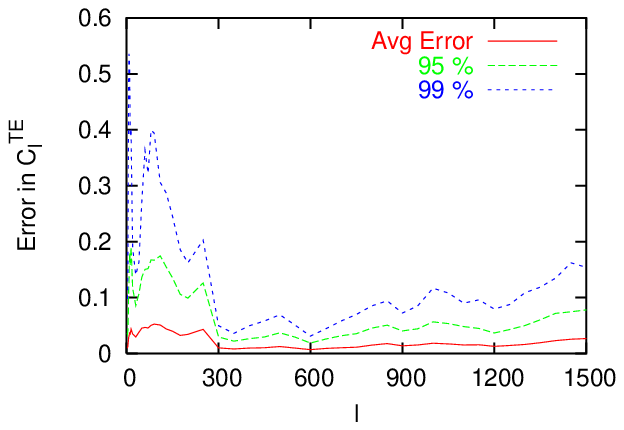}
  \includegraphics[width=2.2 in]{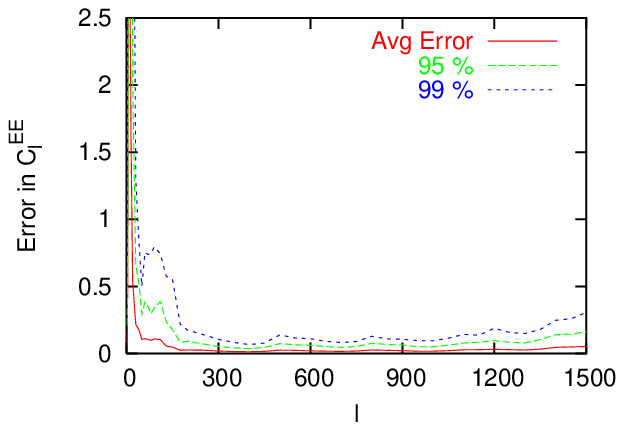}
\caption{\label{fig:spectra} Comparison of the performance of {\sc CosmoNet}
 versus {\sc CAMB} for TT, TE and EE power spectra in 6-parameter flat
 $\Lambda$CDM models. The plots shows the average error together with
 the 95 and 99 percentiles in units of cosmic variance.}
\end{center}
\end{figure*}

It was found that 50 nodes in the hidden layer for the TT spectra
network, 125 nodes for the TE spectra network, 200 nodes for the
EE spectra network, and 50 for the likelihood network, were
sufficient to provide good results. The results of comparing the
{\sc CosmoNet} output with {\sc CAMB} over the testing set are
shown in Fig.~\ref{fig:spectra}.
For all but the very low values of $l$ in the EE spectrum, where
the values of the spectrum and cosmic variance are small, the
average error is about $2-3\%$ of cosmic variance. The
$99^{\hbox{th}}$ percentiles are also comfortably below unit
cosmic variance. A comparison of output of the {\sc CosmoNet} likelihood
network with the WMAP3 likelihood code over the testing set reveals a mean error of roughly 0.2 $\ln$ units
close to the peak.
%

%\begin{figure}
%\begin{center}
%  \includegraphics[width = 5cm, angle = -90]{lhood_errors.eps}
%\caption{\label{fig:lhood} Error in {\sc CosmoNet} likelihood interpolation
%versus the WMAP3 likelihood for the 6-parameter flat
% $\Lambda$CDM models.}
%\end{center}
%\end{figure}

\section{Application to cosmological parameter estimation}
\label{sec:cosmo}

\begin{figure}
\begin{center}
  \includegraphics[width=2.7 in]{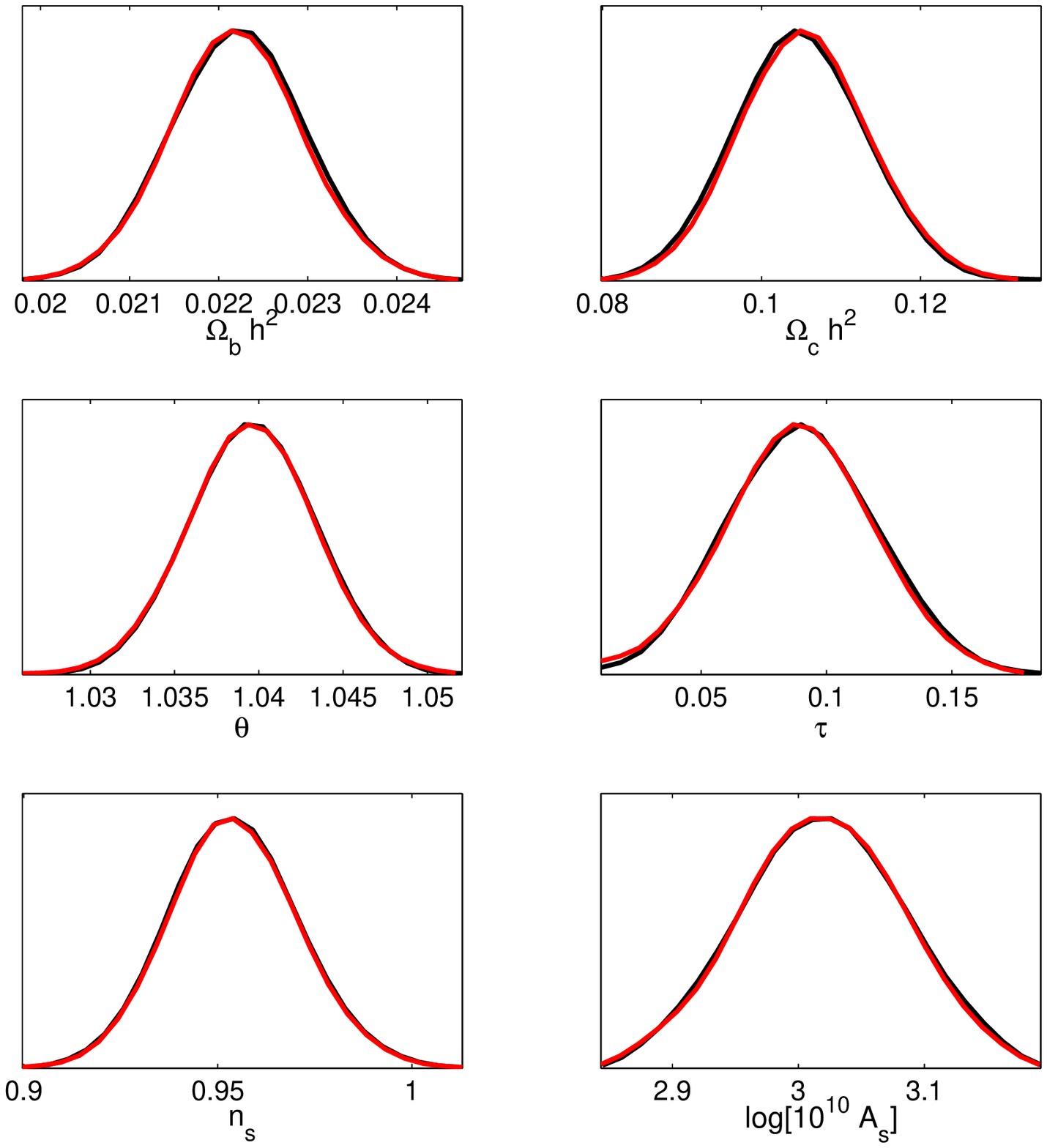}
\caption{\label{fig:wlike} The one-dimensional marginalised
posteriors on the cosmological parameters within the 6-parameter flat
$\Lambda$CDM model comparing: {\sc CAMB}
power-spectra and WMAP3 likelihood (red) with {\sc CosmoNet} power
spectra and WMAP3 likelihood (black).}
\end{center}
\end{figure}

\begin{figure}
\begin{center}
  \includegraphics[width=2.7 in]{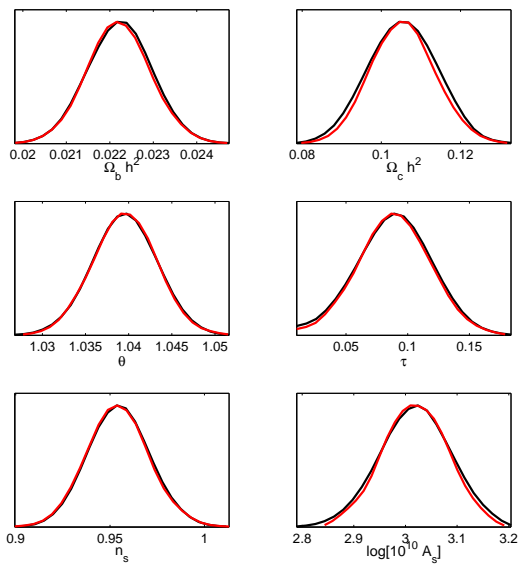}
\caption{\label{fig:wolike} The one-dimensional marginalised
posteriors on the cosmological parameters within the 6-parameter flat
$\Lambda$CDM model comparing: {\sc CAMB}
power-spectra and WMAP3 likelihood (red) with {\sc CosmoNet} likelihoods (black).}
\end{center}
\end{figure}

To illustrate the usefulness of {\sc CosmoNet} in cosmological parameter
estimation we perform an analysis of the WMAP 3-year TT, TE and EE
data using {\sc CosmoMC} in three separate ways: (i) using {\sc CAMB} power spectra
and the WMAP3 likelihood code; (ii) using {\sc CosmoNet} power spectra and the
WMAP3 likelihood code; and (iii) using the {\sc CosmoNet} likelihood. The
resulting marginalised parameter constraints for each method are shown
in Fig.~\ref{fig:wlike} and Fig.~\ref{fig:wolike}, and are clearly very similar.

In each case 4 parallel MCMC chains were run on Intel Itanium 2 processors at 
the COSMOS cluster (SGI Altix 3700) at DAMTP, Cambridge. The wall-clock computational time \footnote{The total CPU time is 4
times longer.} required to gather $\sim 20000$ post
burn-in MCMC samples was $\sim 12$ hours for method (i) (with {\sc CAMB} further parallelised over 3 additional processors per chain,
therefore totalling 16 CPUs), ~8 hours for method (ii) and roughly 35$\pm5$ minutes using the interpolated likelihood with method
(iii). For comparison, a similar run with the {\sc Pico} code took roughly 55$\pm5$ minutes \footnote{Note both method (iii) and {\sc
PICO} were run within the $3\sigma$ training region of both algorithms so {\sc CAMB} was never called.} illustrating that it is now the
remaining sampling calls within {\sc Cosmomc} that  provides the new bottleneck.

\section{Discussion and conclusions}
\label{sec:conc}

We have presented a method for accelerating power spectrum and
likelihood evaluations based on the training of multilayer
perceptron neural networks, which we have shown to be fast, robust
and accurate. Our {\sc CosmoNet} method shares all the advantages of the
Pico algorithm of Fendt \& Wandelt, achieving similar accuracies
on both spectra interpolations and cosmological parameter
constraints, but there are several differences between the two
methods that we believe give {\sc CosmoNet} a number of additional
benefits, which we now discuss.

{\em Simplicity}.  Despite requiring the optimisation of a
highly non-linear multi-dimensional function using MemSys, we
consider the principal advantage of our method to be the relative
simplicity of the trained interpolation for the user to
implement. {\sc CosmoNet} provides a {\em single} simple,
closed-form function for each interpolation over the {\em whole}
of the parameter space under consideration. 

%This means that users
%need not rely on our code to implement the interpolations. Once
%parameters for the networks have been determined, they can be
%released and the feed-forward mapping described in section
%\ref{sec:mlp} can be written easily by anyone in a computer
%language of their choosing. 

%Also, although a $3\sigma$ confidence
%region was used in section \ref{sec:interp}, there are no
%restrictions on the size of parameter space, or the complexity of
%the interpolation. Should the {\sc CAMB} mapping prove more
%complex for larger input spaces, one can simply increase the
%number of hidden nodes in the network. In the {\sc Pico} method,
%however, increasing the size of the parameter space will require
%an increase in the number of clusters into which it is divided.

{\em Memory usage}. A neural network with $N_{\rm in}$ inputs, $N_{\rm
  hid}$ nodes in the hidden layer, and $N_{\rm out}$ outputs has
  $(N_{\rm in} + 1)N_{\rm hid} + (N_{\rm hid} + 1)N_{\rm out} \approx
  N_{\rm hid}N_{\rm out}$ parameters. This is far less than in the
  {\sc Pico} approach, where the use of clustering and individual
  interpolations for each $C_\ell$ requires far more parameters. In
  the case of the flat $\Lambda$CDM example demonstrated in section
  \ref{sec:interp}, we require about $100$ kB of parameter memory for
  all three power spectra and the likelihood, whereas {\sc Pico} would
  require $15$ MB. 
  %The {\sc Pico} memory requirements can be reduced using
  %Karhumen-Lo\`eve compression on the power spectra, but will still be
  %at least 2 orders of magnitude greater than our neural network
  %approach. The Karhumen-Lo\`eve compression is neither required nor
  %appropriate with the neural network method. Karhumen-Lo\`eve
  %essentially reduces the dimension of the output space of the
  %interpolation through finding orthogonal linear combinations of
  %spectra components with greatest variance. As {\sc Pico} has a separate
  %polynomial interpolation for each $C_\ell$ this is useful. However,
  %the maximally varying components of the outputs are found in the
  %activations of the nodes in the hidden layer in the neural
  %network. The map from each hidden node to the output layer {\it is}
  %linear (up to additive constant) and will generate similar, albeit
  %non-orthogonal, components to those found by Karhumen-Lo\`eve.
  While the memory requirements of {\sc Pico} will increase with the number
  of cosmological parameters, this should make little difference to
  the memory requirements of our method, as we have found the required number 
  of nodes in the hidden layer does not increase beyond a factor of 2 
  for the 11 parameter non-flat model parameterised by $\Omega_b h^2$, $\Omega_c h^2$, 
  $\Omega_k$,
  $\theta$, 
  $\tau$, massive neutrino fraction $f_{\nu}$, varying equation of state of dark energy $w$, scalar
  perturbation amplitude and spectral index $A_s$, $n_s$ and tensor modes with amplitude ratio and
  spectral index $R$, $n_t$ (results to appear in a forthcoming
  paper), thus representing a linear rise.

{\em Speed}. The number of calculations required to perform the
feed-forward network mapping is $2 N_{\rm in} \ N_{\rm hid} \ + \
2 N_{\rm hid} \ N_{\rm out} \approx 2 N_{\rm hid} N_{\rm out} \ $.
In the example presented in section \ref{sec:interp} the
calculation of the 50 uninterpolated $C_\ell$ values for each
spectrum required $\sim$ 120 microseconds, whereas each WMAP3
likelihood took $\sim$ 10 microseconds; this is $\sim 25$ times
faster than {\sc Pico} in performing the interpolation. Moreover,
the CPU requirements of the {\sc Pico} interpolation scales as $(
\frac{N_{\rm in}}{p})^p$, for {\sc Pico} $p$ is 4.

{\em Ease of training}. Training using the {\sc MemSys} package is
almost totally automated and relatively quick. In fact the
bottleneck in providing appropriate network weights for more
complex cosmological models is the calculation of the training
data using the {\sc CAMB} package. The networks used in section
\ref{sec:interp} took around 100 hours of training (on a standard
PC workstation). Additionally, MemSys training time scales
linearly with the number of network nodes and again linearly with
the number of entries in the training set. However, networks with
an accuracy of roughly $2-3$ times worse, that are sufficient to
provide good parameter constraints can be trained in under an hour
using just 1000 training samples. These simpler neural
networks did not require the l-splitting performed in section
\ref{sec:interp} and had only 50 nodes in the hidden layer.

\medskip

%The homepage for the {\sc CosmoNet} package is located at {\tt
%www.mrao.cam.ac.uk/software/cosmonet} and contains the
%network parameters for the interpolation presented in
%Section~\ref{sec:interp}.
% which can easily be used to write an
%interpolation routine. We also provide a simple interpolation routine that
%interfaces with {\sc CosmoMC} for those users who do not wish to write their
%own. We plan to release network parameters for a wider range of
%(non-flat) cosmological models shortly. For those wishing to use the
%neural networks for other non-standard models, we will produce
%appropriate network parameters on request provided we are supplied
%with suitable training data. In the longer term, subject to
%negotiation with the developers of {\sc MemSys}, we hope to release the
%network training code, so that users may train up their own networks.

%The homepage for the {\sc CosmoNet} package is located at {\tt
%www.mrao.cam.ac.uk/software/cosmonet} which contains the network parameters used for the
%interpolations and routines that interface with {\sc CosmoMC}. 

\subsection*{ACKNOWLEDGMENTS}
TA acknowledges a studentship from EPSRC. MB was supported by a Benefactors Scholarship
at St. John's College, Cambridge and an Isaac Newton studentship. This work was
conducted in cooperation with SGI/Intel utilising the Altix 3700 supercomputer
at DAMTP Cambridge supported by HEFCE and PPARC. We thank S. Rankin and V.
Treviso for their assistance.

\bsp % ``This paper has been produced using the ...''
\label{lastpage}
\end{document}